\documentclass[%
reprint,
superscriptaddress,
 amsmath,amssymb,
prfluids
longbibliography,
]{revtex4-2}

\usepackage{romannum}
\usepackage{graphicx}% Include figure files
\usepackage{dcolumn}% Align table columns on decimal point
\usepackage{siunitx}
\usepackage{hyperref}

\usepackage{xcolor}
\definecolor{myred}{RGB}{255,3,0}
\definecolor{myblue}{RGB}{31,60,255}
\definecolor{myblue2}{RGB}{34,123,255}
\definecolor{myorange}{RGB}{226,48,0}
\definecolor{mygreen}{RGB}{21,100,0}
\definecolor{mygreen2}{RGB}{118,172,66}

\usepackage{tikz}
\usetikzlibrary{shapes.geometric,shapes.symbols}

\begin{document}

\title{\large{Hysteresis in the freeze-thaw cycle of emulsions and suspensions}}

\author{Wilfried Raffi}
\thanks{These two authors contributed equally}
\email[]{wilfried.raffi@ens.psl.eu}
\affiliation{Physics of Fluids group, Max Planck Center Twente for Complex Fluid Dynamics, Department of Science and Technology, Mesa+ Institute and J. M. Burgers Center for Fluid Dynamics, University of Twente, P.O. Box 217, 7500 AE Enschede, The Netherlands}

\author{Jochem G. Meijer}%
\thanks{These two authors contributed equally}
\email[]{jgmeijer@uchicago.edu}
\affiliation{Physics of Fluids group, Max Planck Center Twente for Complex Fluid Dynamics, Department of Science and Technology, Mesa+ Institute and J. M. Burgers Center for Fluid Dynamics, University of Twente, P.O. Box 217, 7500 AE Enschede, The Netherlands}

\author{Detlef Lohse}%
\email[]{d.lohse@utwente.nl}
\affiliation{Physics of Fluids group, Max Planck Center Twente for Complex Fluid Dynamics, Department of Science and Technology, Mesa+ Institute and J. M. Burgers Center for Fluid Dynamics, University of Twente, P.O. Box 217, 7500 AE Enschede, The Netherlands}
\affiliation{Max Planck Institute for Dynamics and Self-Organization, Am Fassberg 17, 37077 G\"ottingen, Germany}

\begin{abstract}
Freeze-thaw cycles can be regularly observed in nature in water and are essential in industry and science. 
Objects present in the medium will interact with either an advancing solidification front during freezing or a retracting solidification front, \textit{i.e.}, an advancing melting front, during thawing.
It is well known that objects show complex behaviours when interacting with the advancing solidification front, but the extent to which they are displaced during the retraction of the solid-liquid interface is less well understood.
To study potential hysteresis effects during freeze-thaw cycles, we exploit experimental model systems of oil-in-water emulsions and polystyrene (PS) particle suspensions, in which a water-ice solidification front advances and retracts over an individual immiscible (and deformable) oil droplet or over a solid PS particle. 
We record several interesting hysteresis effects, resulting in non-zero relative displacements of the objects between freezing and thawing.
PS particles tend to migrate further and further away from their initial position, whereas oil droplets tend to return to their starting positions during thawing.
We rationalize our experimental findings by comparing them to our prior theoretical model of Meijer, Bertin \& Lohse, Phys. Rev. Fluids (2025) \cite{meijer2024frozen}, yielding a qualitatively good agreement.
Additionally, we look into the reversibility of how the droplet deforms and re-shapes throughout one freeze-thaw cycle, which will turn out to be remarkably robust.
\end{abstract}

\date{\today}

\maketitle
\onecolumngrid

\section{Introduction}

Liquids exposed to thermal gradients can freeze, forming a solid-liquid interface (solidification front) that will propagate in the direction of the applied thermal gradient. 
If seeded with solid particles, droplets or bubbles, these objects will interact with the advancing solidification front.
They can either be pushed away by the front, remaining submerged in the liquid, or can (eventually) pass through the interface and become trapped in the solid \cite{korber1985interaction,shangguan1992analytical,lipp1993engulfment,dedovets2018five, tao2016steady,garvin2007multiscale,park2006encapsulation, tyagi2021multiple, rempel2001particle,rempel1999interaction,tyagi2020objects,meijer2024frozen}.
Immersed objects that are soft and deformable are additionally subjected to stresses during the encapsulation process, leading to their deformation \cite{tyagi2022solute,meijer2023thin}, or even sudden topological transitions \cite{meijer2024freezing}.
In the case of gas bubbles, mass transfer adds an additional layer of complexity, resulting in a variety in the shapes and sizes of the bubbles captured in ice \cite{carte1961air,bari1974nucleation,wei2000shape,wei2002analytical, wei2004growths,thievenaz2024universal,meijer2024enhanced}.

Understanding these interactions and being able to manipulate them in a controlled manner is of great interest for many industrial applications \cite{deville2017freezing}, ranging from templating
directionally porous materials \cite{deville2008freeze,deville2022complex}, to proper cryo-preservation procedures for food \cite{amit2017review} and biological tissues \cite{bronstein1981rejection,korber1988phenomena,muldrew2004water}.
As a result, much attention has been given to studying these systems in detail under various freezing conditions.
While many freezing procedures aim to store and preserve samples for extended periods of time, thawing, however, is often inevitable. 
How the thawing process affects the arrangement of submerged particles, and how this differs from the freezing process, remains an open question.

In this paper, we therefore aim to illuminate how different types of hystereses can occur during a single freeze-thaw cycle in oil-in-water emulsions and polystyrene (PS) particle suspensions. 
This is achieved through well-controlled, uni-directional freezing/thawing experiments on sub-millimetre, single-particle systems.
We report our experimental methods in section \ref{sec:Exp}, section \ref{sec:Results} reports our observations, section \ref{sec:Rev} addresses the reversibility and in section \ref{sec:Model} we compare our results to predictions of our prior theoretical model of \cite{meijer2024frozen}.
We end with suggesting a possible extension of the model in order to rationalize some of our experimental findings, as well as a conclusion and an outlook (section \ref{sec:Conclusion}).

\begin{figure}  % Figure over entire width
\includegraphics[width=\textwidth]{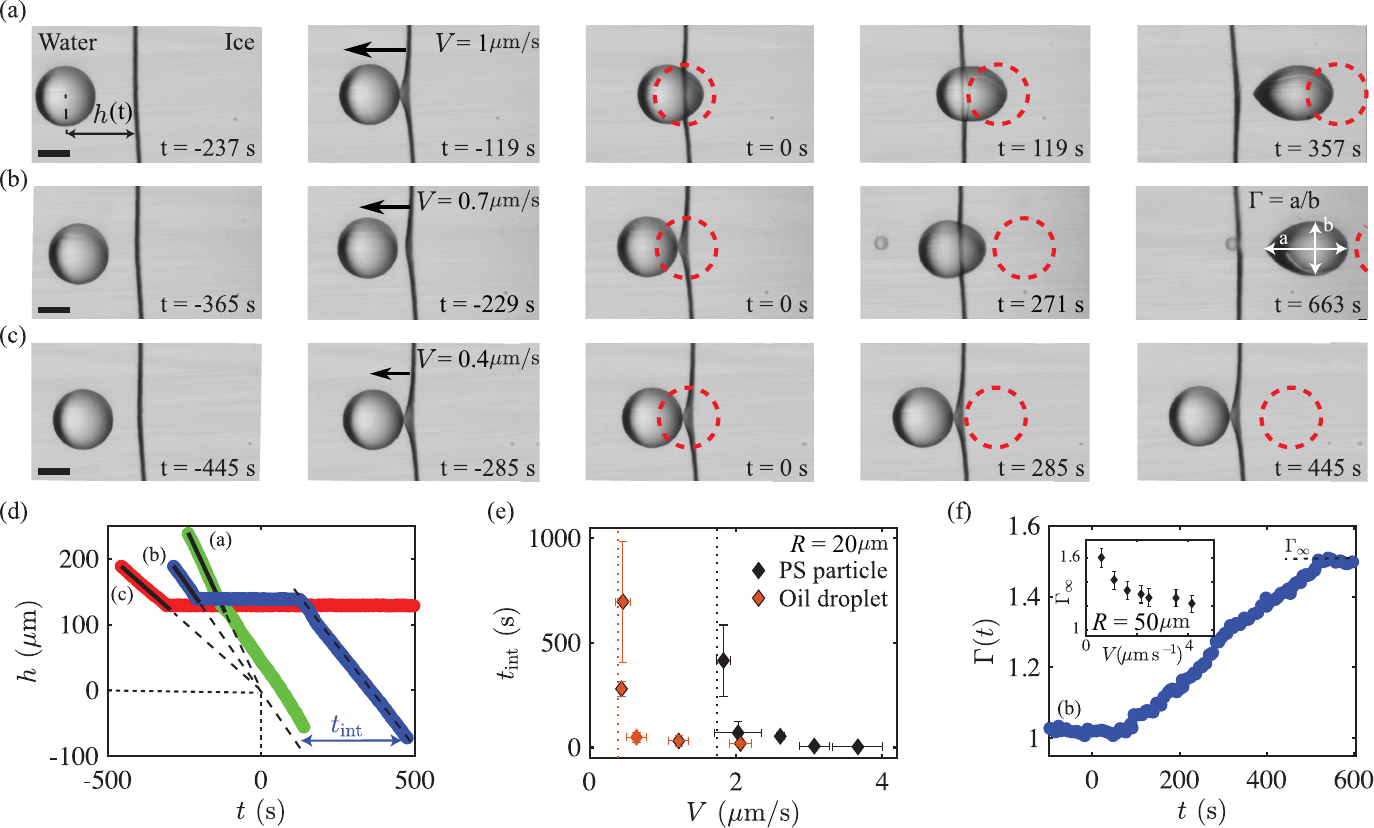}
\centering
\caption{Interactions during freezing. (a)-(c) Interaction between a silicone oil droplet of size $R \approx \SI{105}{\micro \meter}$ with a water-ice solidification front advancing at different velocities $V$, \textit{i.e.},  (a) $V \approx \SI{1}{\micro \meter \per \second}$, (b) $V \approx \SI{0.7}{\micro \meter \per \second}$,  and (c) $V \approx \SI{0.4}{\micro \meter \per \second}$ (see Supplementary Movies 1-3 \cite{suppl}).  Images are taken in the frame of reference of the moving front. Depending on the rate of approach, the droplet (a) does barely interact with the front and is rapidly engulfed into the ice, (b) interacts with the front for a certain amount of time $t_{\mathrm{int}}$ before being engulfed, or (c) is repelled by the ice indefinitely. During the encapsulation into the ice the droplet deforms \cite{tyagi2022solute,meijer2023thin}.  The red contours indicate the position where the droplet would have been if the particle-front interaction would have been absent, assuming linear drop/particle motion.  Time $t=0$ is defined as the moment in time the center of the particle would have reached the undeformed front. Scale-bars are $\SI{100}{\micro \meter}$.
(d) Particle-front distance $h(t)$ as a function of time for the three representative cases, \textit{i.e.}, (a) fast engulfment, (b) intermediate rejection, and (c) indefinite rejection.   
(e) Particle-front interaction time $t_{\mathrm{int}}$ as a function of advancing velocity $V$ for both oil droplets and polystyrene (PS) particles with $R \approx \SI{20}{\micro \meter}$. 
As $V$ approaches a certain critical value $V_{\mathrm{crit}}$ (dotted lines) the interaction time rapidly increases \cite{meijer2024frozen}. 
(f) Aspect ratio $\Gamma(t)$ as a function of time quantifying the deformation dynamics during encapsulation of the droplet corresponding to (b).  The inset shows the final extent of the deformation $\Gamma_{\infty}$, averaged over at least three droplets, as a function of $V$ for oil droplets with $R \approx \SI{50}{\micro \meter}$. \cite{tyagi2022solute,meijer2023thin}}
\label{fig:1}
\end{figure}

\section{Experimental methods}
\label{sec:Exp}

In order to study the fundamentals of the interaction between a particle and an advancing/retracting solidification front, we make use of horizontal, uni-directional freezing/thawing.
Details of the experimental set-up and procedures are provided in earlier work \cite{meijer2024frozen}.
In short, we expose a $\SI{200}{\micro \meter}$ thick Hele-Shaw cell (Ibidi, $\mu$-Slide I Luer) to a fixed thermal gradient. 
The Hele-Shaw cell is filled with our working liquids, \textit{i.e.}, either an oil-in-water emulsion or a PS particle suspension, where the bulk phase consists of Milli-Q water.
For the former we use 5 cSt silicone oil (Sigma-Aldrich, Germany) and a small amount (0.01 vol\% after dilution \cite{meijer2024frozen}) of surfactant TWEEN-80 (Sigma-Aldrich, Germany) to stabilise the emulsion.
The emulsions are prepared in a way that yields polydisperse droplet sizes.
For the latter,  PS particles of two different sizes ($R \approx \SI{20}{\micro \meter}$ and $R \approx \SI{70}{\micro \meter}$) are employed (Microbead Dynoseeds TS40 and TS140).
Since we are interested in the interaction between a freezing/thawing front with an \textit{individual and isolated} spherical object we ensure that the emulsions and suspensions are sufficiently dilute.

A fixed thermal gradient over the Hele-Shaw cell, which rests on two copper blocks spaced roughly $\SI{3}{\mm}$ apart, is achieved by cooling down one side to $\left(-15\pm0.2\right)^{\circ}$C.
The other side is kept at a constant temperature of $\left(18\pm0.2\right)^{\circ}$C, resulting in a thermal gradient in the order of $G \approx \SI{1e4}{\kelvin \per \meter}$.
The temperatures on both sides are constantly monitored using thermocouples.
The entire system is placed inside a humidity control box to prevent fog and frost formation that would obscure the view.

Once a stable thermal gradient is reached, we trigger solidification on the cold side, leading to the formation of a planar solid-liquid interface, parallel to the temperature gradient, that will enter our field of view and will eventually reach a stable equilibrium position. 
To study the interaction of particles with an advancing solidification front, we then move the Hele-Shaw cell over the copper blocks towards the cold side at a constant velocity $V$ using a high-precision linear actuator (Physik Instrumente, M-230.25).
The position of the planar front remains fixed in space throughout the experiments, while the ice keeps on growing.
The approach velocity of the object towards the solidification front can then be controlled with an uncertainty of around 5\% for velocities of the order of $V \approx \SI{0.1}{\micro \meter \per \second}$, and of around 2\% for $V \approx \SI{1}{\micro \meter \per \second}$. 
We note that when viewed from the frame of reference of the Hele-Shaw cell, this is equivalent to a solidification front sweeping through the emulsion/suspension at the same velocity $V$.
Once the solidification front has swept over the particles we initiate the thawing process. For this, $V$ is slowly decreased to zero. Once the position of the front is equilibrated again, a retracting, planar solidification front is then achieved by moving the Hele-Shaw cell in the opposite direction, \textit{i.e.}, towards the warm side, at a constant velocity $V$. 
The rate at which the front advances or retracts can be altered independently, and it does not have to be the same. 
This protocol allows us to study a freeze-thaw cycle on the same particle with the same or varying freezing and thawing rates.

Finally, snapshots are taken from above at regular intervals in the region between the two copper plates where the solidification front is located.
A Nikon D850 camera with a long working distance lens is used for this purpose.
The sample is illuminated with cold-LED back-lighting to avoid localised heating.

\section{Experimental results on freezing and thawing}
\label{sec:Results}

\begin{figure}  % Figure over entire width
\includegraphics[width=\textwidth]{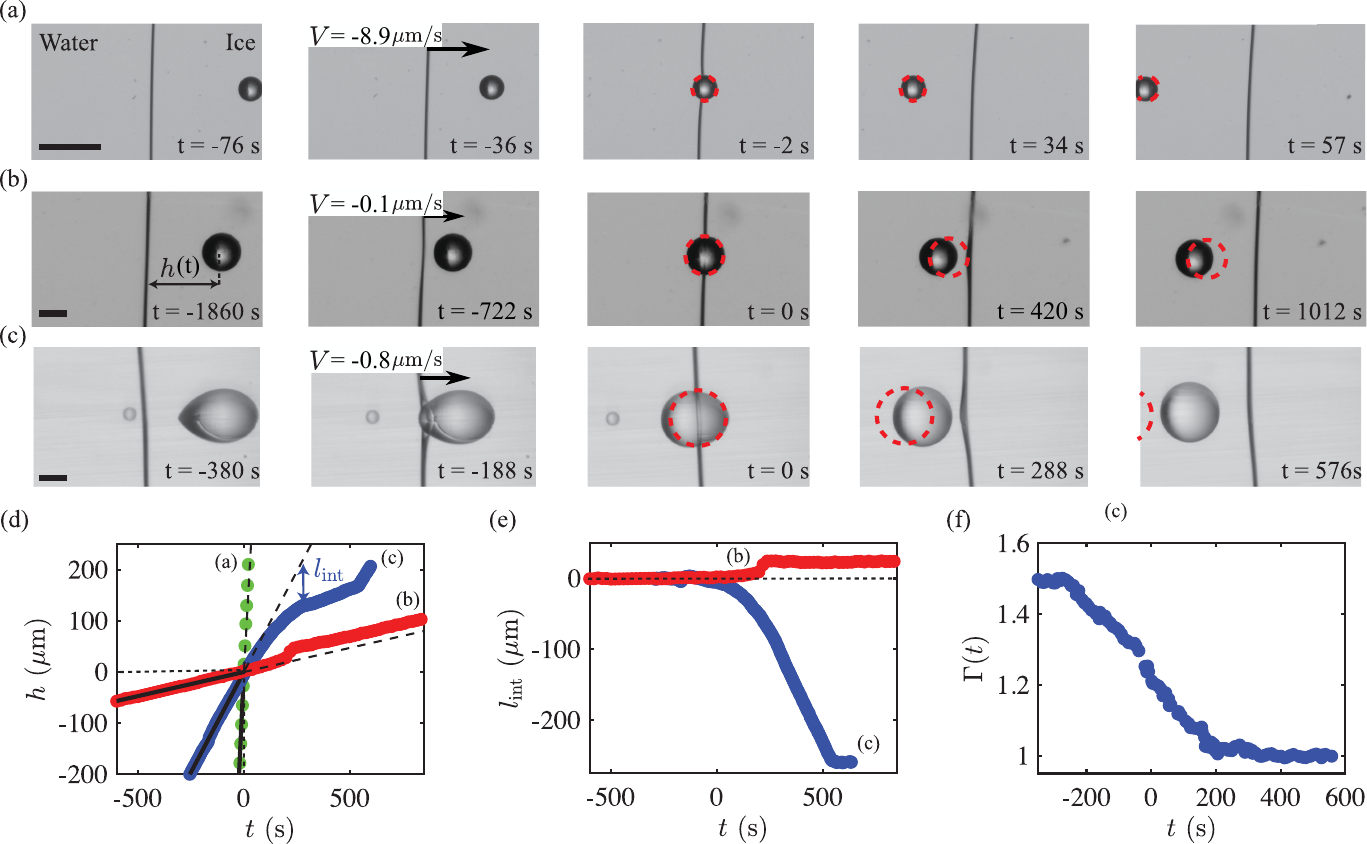}
\centering
\caption{Interactions during thawing. (a)-(c) Interaction between a PS particle of size (a) $R \approx \SI{20}{\micro \meter}$ and (b) $R \approx \SI{70}{\micro \meter}$, and a (c) silicone oil droplet of size $R \approx \SI{105}{\micro \meter}$ with a water-ice solidification front retracting at different velocities $V$, \textit{i.e.},  (a) $V \approx \SI{-8.9}{\micro \meter \per \second}$, (b) $V \approx \SI{-0.1}{\micro \meter \per \second}$,  and (c) $V \approx \SI{-0.8}{\micro \meter \per \second}$ (see Supplementary Movies 4-6 \cite{suppl}).  Images are taken in the frame of reference of the moving front. Depending on the rate of retraction and the type of particle, the object (a) does barely interact with the front and is rapidly expelled by the ice, (b) experiences an additional push by the retracting front, leading to a sudden displacement in the direction opposite of the motion of the front, or (c) is being held back by the front for a certain amount of time. 
We note that for (b) and (c) it is the type of object that dictates the type of interaction during thawing, rather than its size (see Suppl.\,Mat.\,III \cite{suppl}).
During the extraction out of the ice the droplet regains its spherical shape.  The red contours indicate the position where the droplet would have been if the particle-front interaction would have been absent.  Time $t=0$ is defined as the moment in time the center of the particle would have reached the undeformed front. The denoted scale-bars are $\SI{100}{\micro \meter}$.
(d) Particle-front distance $h(t)$ as a function of time for the three representative cases, \textit{i.e.}, (a) fast extraction without interaction, (b) sudden additional displacement opposite the direction of motion of the front for the PS particle, and (c) retardation of the motion away from the front for the oil droplet.   
(e) Particle-front interaction length $l_{\mathrm{int}}$ as a function of time for the (b) PS particle and (c) oil droplet, highlighting the difference in particle displacement during thawing. 
(f) Aspect ratio $\Gamma(t)$ as a function of time, quantifying the re-formation dynamics during extraction of the droplet corresponding to (c). }
\label{fig:2}
\end{figure}

We will now present a detailed discussion of our experimental results.  We will quantify the interaction of the droplet/particle with the solidification front through the experimentally most accessible variable, namely the particle-front distance $h(t)$ (see first panel Fig.\,\ref{fig:1}\,(a)). It is defined as the instantaneous distance of the center of the particle and the mid-plane of the undeformed front.

\subsection{Freezing}
During the freezing process the object approaches the planar solidification front at a constant velocity $V$ (or vice versa). 
As has been well established in prior studies, three distinct regimes can be observed for both oil droplets and solid particles, depending on their rate of approach \cite{rempel1999interaction, rempel2001particle,park2006encapsulation, tyagi2022solute,meijer2024frozen}.
If the approach occurs rapidly, well above a certain critical value $V_{\mathrm{crit}}$, the object barely interacts with the solidification front and is quickly incorporated into the ice (see Fig.\,\ref{fig:1}\,(a)\,\&\,(d) for the case of a slightly confined oil droplet of size $R \approx \SI{105}{\micro \meter}$).
Oppositely, if $V < V_{\mathrm{crit}}$ the object is indefinitely repelled by the front and will never get trapped (see Fig.\,\ref{fig:1}\,(b)\,\&\,(d)), eventually leading to the creation of pure ice.
Lastly, at values slightly above $V_{\mathrm{crit}}$, the object interacts with the solidification front for a certain amount of time $t_{\mathrm{int}}$ before suddenly entering the ice (see Fig.\,\ref{fig:1}\,(c)\,\&\,(d)).
The precise definition of $t_{\mathrm{int}}$ for the case shown in Fig.\,\ref{fig:1}\,(b) is depicted in Fig.\,\ref{fig:1}\,(d) and taken form the experimentally measured profile of $h(t)$.
During this interaction the droplet is displaced over a distance $l_{\mathrm{int}}$ in the direction of motion of the front.
The time the particle spends at the solid-liquid interface, and the extent of the experienced displacement, thus depends on $V$ (see Fig.\,\ref{fig:1}\,(e)), but also on the size of the particle, the strength of the applied thermal gradient, as well as other (chemical) properties of the particle-liquid-solid system, which all dictate the precise value of $V_{\mathrm{crit}}$ \cite{rempel1999interaction, rempel2001particle,park2006encapsulation}.
When performing freezing experiments for oil droplets and PS particles both with $R \approx \SI{20}{\micro \meter}$, we find $V_{\mathrm{crit, oil}} \approx \SI{0.4}{\micro \meter \per \second}$ and $V_{\mathrm{crit, PS}} \approx \SI{1.8}{\micro \meter \per \second}$, respectively (see Fig.\,\ref{fig:1}\,(e)).
For larger sizes, these values might shift upwards \cite{meijer2023thin} but they will remain significantly different for both types of systems.

We turn our focus to the representative case depicted in Fig.\,\ref{fig:1}\,(b) of the oil droplet that interacts with the front.
One other remarkable observation is that already before the droplet makes contact with the front, its mere presence can cause the initially planar front to bend (see second panel Fig.\,\ref{fig:1}\,(b)). 
It has been established that this feature arises due to the thermal conductivity mismatch between the particle and the surrounding melt causing the isotherms around the particle (and hence the solid-liquid interface) to bend \cite{shangguan1992analytical,tyagi2020objects,van2024deforming}.
For the two systems studied here the thermal conductivity mismatch between that of the particle ($k_p$) and that of the surrounding liquid ($k_l$) are $k_p/k_l = 0.05$ and $k_p/k_l = 0.2$ for the PS particles and oil droplets in water, respectively.
Surprisingly, this deflection can be altered and even reversed when introducing thermo-capillary (\textit{i.e.}, thermal Marangoni) flows at the free surface of the droplet, triggered by the applied thermal gradient \cite{van2024deforming}.
Once the particle-front interaction has taken place the droplet enters the ice.
During its encapsulation the droplet is subjected to mechanical stresses, causing it to compress in the direction parallel to the front and to elongate in the perpendicular direction, eventually assuming a pointy, tear-like shape that persists throughout the duration of the experiments, which can take up to several hours \cite{meijer2023thin}.
To quantify the deformation dynamics we introduce an aspect ratio $\Gamma(t)$ (see last panel Fig.\,\ref{fig:1}\,(b)).
The evolution for this case is depicted in Fig.\,\ref{fig:1}\,(f), where the droplet starts off spherical ($\Gamma = 1$), before deforming through two distinct regimes towards a final value $\Gamma_{\infty}$.
By performing consecutive freeze-thaw cycles on the same droplet with $R = \SI{50}{\micro \m}$ at increasing freezing/thawing velocities, we retrieve that the final extent of the droplet deformation $\Gamma_{\infty}$ does depend on $V$, where a faster approach leads to less deformation (see inset of Fig.\,\ref{fig:1}\,(f)). For these cases we have also analysed their temporal evolution and find that the characteristic kink in the curve corresponds to the moment in time that $h=0$ \cite{meijer2023thin}, \textit{i.e.},  the center of the particle overlaps with the mid-plane of the undeformed front. In addition, we find that the different profiles of $\Gamma(t)$ at different freezing velocities $V$ collapse after normalisation (see Suppl.\,Mat.\,II \cite{suppl}).
Needless to say that solid particles are rigid enough to remain spherical and do not deform during encapsulation.

\subsection{Thawing}

Now we shift towards the main findings of our paper and reverse the process. 
We ensure that our sample moves in the opposite direction, causing the ice to slowly melt and the solid-liquid interface to retract in a controllable fashion. 
Similar to rapid freezing, rapid thawing does not allow for the particle or droplet to interact with the retracting front and no further displacement is experienced by the object as it re-enters the melt (see Fig.\,\ref{fig:2}\,(a)\&\,(d) for a PS particle with $R \approx \SI{20}{\micro \meter}$).

Slowing down the thawing process, however, leads to the observation of interesting interactions.
Whereas for freezing a critical velocity exists below which encapsulation into the ice does not occur, there is no such limitation for thawing.  
Independent on how slow the retraction might be, the object will always end up in the melt.
Alternatively, in the case of PS particles with $R \approx \SI{70}{\micro \meter}$, and for a sufficiently slow retracting solidification front (here $V \approx \SI{-0.1}{\micro \meter \per \second}$), the particle remarkably experiences an additional push by the retracting front, leading to a sudden displacement in the opposite direction of the motion of the front (see Fig.\,\ref{fig:2}\,(b)\&\,(d)).
This particular feature becomes more apparent when looking into the particle displacement $l_{\mathrm{int}}$ during thawing, see the red data in Fig.\,\ref{fig:2}\,(e), where a sudden positive displacement is clearly visible.
We find that such additional displacement becomes smaller when the rate of retraction is increased (see Suppl.\,Mat.\,III \cite{suppl}).\\
Lastly, for the case of the oil droplet, yet another different observation can be made. 
For retracting velocities with values close to $V_{\mathrm{crit}}$ the motion of the droplet away from the retracting front seems to be retarded as the droplet enters the melt (see  Fig.\,\ref{fig:2}\,(c)\&\,(d)).
In other words, the droplet slows down, even after the droplet has exited the ice, and is held back by the retracting front for a specific period of time, before finally moving away again at constant velocity $V$, see Fig.\,\ref{fig:2}\,(d) and Supplementary Movie 6 \cite{suppl}.
The extent of the displacement of the droplet in the direction of the moving front can be quantified once more using $l_{\mathrm{int}}$.
The data points shown in blue in Fig.\,\ref{fig:2}\,(e) clearly show the significance of this displacement.
We find that the smaller oil droplet with $R = \SI{70}{\micro \meter}$ is less susceptible to the retracting front and that the extent of the droplet displacement quickly diminishes at faster thawing rates (see Suppl.\,Mat.\,III \cite{suppl}).

As the deformed droplet re-enters the melt it will regain its spherical shape.
In the same fashion as above, we determine the instantaneous aspect ratio $\Gamma(t)$ of the droplet in order to quantify the re-formation dynamics. 
Fig.\,\ref{fig:2}\,(f) depicts a typical curve corresponding to Fig.\,\ref{fig:2}\,(c), where the droplet regains its spherical shape ($\Gamma = 1$), with an apparently similar but reversed dynamics as during freezing.  

\section{Reversibility}
\label{sec:Rev}

\begin{figure}[b!]
\includegraphics[width=\columnwidth]{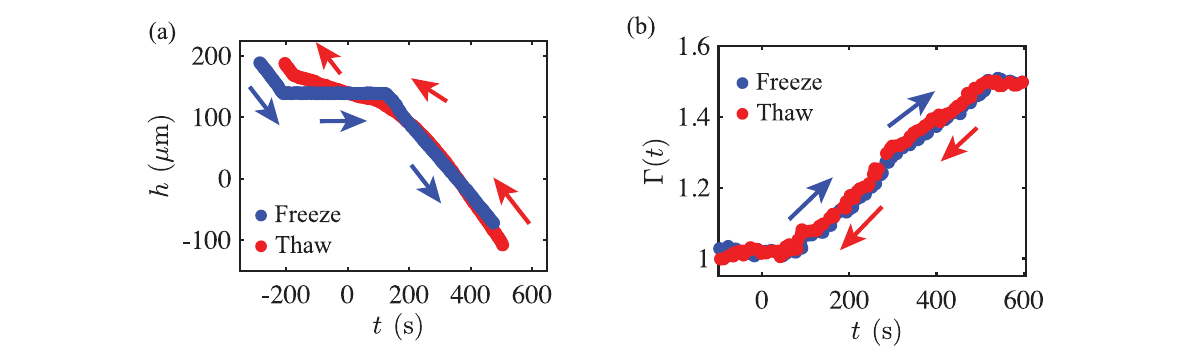}
\centering
\caption{Reversibility of the freeze-thaw cycle for oil droplets. (a) Particle-front distance $h(t)$ as a function of time for freezing (blue, see Fig.\,\ref{fig:1}\,(b)\,\&\,(d)) and thawing (red, see Fig.\,\ref{fig:2}\,(c)\,\&\,(d)).  The time for the thawing curve (red) has been inverted and then shifted with $t_{\mathrm{int}}$ to let both curves overlap to highlight the effect of hysteresis for this particular case.
(b) Deformation (blue, see Fig.\,\ref{fig:1}\,(f)) and re-formation (red, see Fig.\,\ref{fig:2}\,(f)) dynamics of an oil droplet with $R = \SI{105}{\micro \meter}$ during one freeze-thaw cycle, with $V \approx \SI{0.7}{\micro \meter \per \second}$ for freezing and $V \approx \SI{0.8}{\micro \meter \per \second}$ for thawing. The thawing time has again been inverted and shifted to ensure that the start of the deformation and the end of the re-formation match. }
\label{fig:3}
\end{figure}

To highlight discrepancies and similarities in the dynamic response of the droplet during one freeze-thaw cycle we overlap the experimentally obtained results for both the evolution of the particle-front distance $h(t)$, as well as the deformation parameter $\Gamma(t)$.
We choose the most representative case corresponding to Fig.\,\ref{fig:1}\,(b) and Fig.\,\ref{fig:2}\,(c) for the freezing and thawing process of the same droplet, respectively.
Here, the rate of freezing and thawing are comparable and above $V_{\mathrm{crit}}$.
More precisely, $V \approx \SI{0.7}{\micro \meter \per \second}$ for freezing and $V \approx \SI{0.8}{\micro \meter \per \second}$ for thawing.
To ensure overlap between the freezing and thawing curves of the particle-front distance $h(t)$, the thawing curve (see Fig.\,\ref{fig:2}\,(d)) has been inverted in time and shifted with $t_{\mathrm{int}}$.
The obtained result is shown in Fig.\,\ref{fig:3}\,(a).
The figure reveals the difference in the droplet's response to both processes and the occurrence of hysteresis. 
How this affects the overall displacement of the droplet is discussed in the following section.

For the evolution of the droplet deformation we again invert and shift time for the thawing curve (see Fig.\,\ref{fig:2}\,(f)) to ensure overlap with that of freezing, see Fig.\,\ref{fig:3}\,(b).
Remarkably, we find that this process is perfectly reversible, where the dynamics during deformation and re-formation seem identical (also see Suppl.\,Mat.\,II \cite{suppl}).

\section{Overall particle displacement during one freeze-thaw cycle}
\label{sec:Model}

As shown above, the particles and droplets can experience nontrivial displacements when undergoing a freeze-thaw cycle.
In this section, we briefly revisit the main experimental findings to rationalise these observations through the current theoretical understandings.

\subsection{Experimental observations}

The most convenient way to depict the overall particle displacement during one freeze-thaw cycle is through the earlier introduced particle-front interaction length $l_{\mathrm{int}}$.
To recall, this parameter is defined as the displacement the particle experiences as it interacts with the front for a certain amount of time, $t_{\mathrm{int}}$, see Fig.\,\ref{fig:1}\,\&\,\ref{fig:2}\,(d) for their respective definitions, and specifically reads
\begin{equation}
l_{\mathrm{int}} = \int_0^\infty u[h(t)] \mathrm{d}t.
\label{eq:lint}
\end{equation} 
Here, $u(t)$ is the particle velocity and the variation in the particle-front distance $h(t)$ is then given by
\begin{equation}
\frac{\mathrm{d}h}{\mathrm{d}t} = u(t) - V.
\label{eq:dhdt}
\end{equation}
If $u(t) = V$, the particle-front distance $h(t)$ would remain constant. 
Figure\,\ref{fig:4} shows the evolution of this parameter as a function of time during both freezing (blue) and thawing (red) for an (a) PS particle and (b) oil droplet. 
Whereas the PS particle is pushed further away from its initial position, even when the front retracts, the oil droplets tend to return to its initial position, hence experiencing barely any overall displacement.
It should be emphasized that the extent of the experienced displacements during both freezing and thawing is extremely sensitive to various environmental conditions, such as the value of $V$ and its intrinsic fluctuations, and hence might vary significantly from case to case.
Nonetheless, the obvious difference in the dynamic response of the PS particle as compared to that of the the oil droplet remains apparent. A possible reasoning is provided below, after having introduced the current standards of the theoretical modelling.

\subsection{Theoretical modelling}

Theoretical models of the interaction between a spherical object and an advancing solidification front have been developed over the years \cite{shangguan1992analytical, tao2016steady,garvin2007multiscale,park2006encapsulation, rempel2001particle,rempel1999interaction} and are able to capture the relevant physics in order to at least qualitatively match experimental observations during freezing. 
The question raised here is: what is the performance of these models when, after having modelled the initial part of the interactions, we let the front retract at a certain set velocity?

\begin{figure}[t!]
\includegraphics[width=\columnwidth]{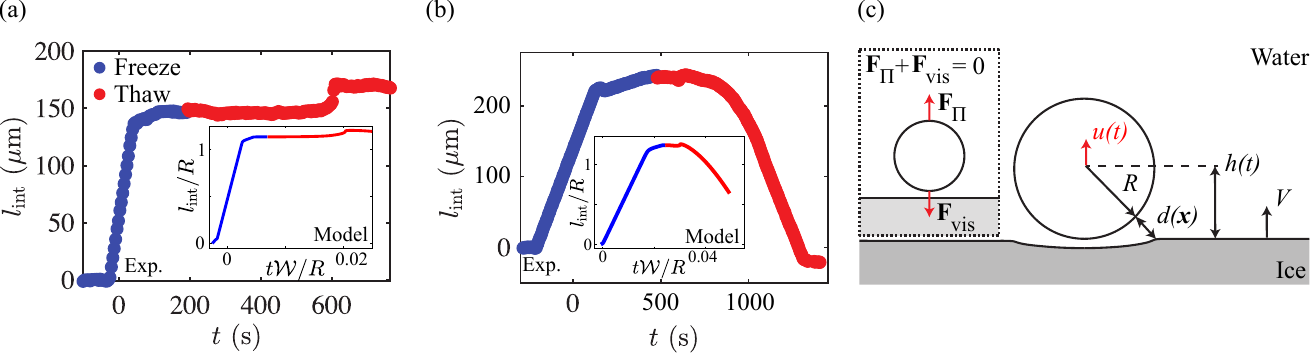}
\centering
\caption{Overall particle displacement during one freeze-thaw cycle. (a) Particle-front interaction length $l_{\mathrm{int}}$ of the PS particle during both freezing (blue, not previously shown) and thawing (red, see Fig.\,\ref{fig:2}\,(b)\,\&\,(e)) as a function of time. The thawing time has been shifted for the curve to become continuous. The inset shows the theoretical prediction of the slightly adjusted model of Ref.\cite{meijer2024frozen} including volume expansion during phase change.
(b) Particle-front interaction length $l_{\mathrm{int}}$ of the oil droplet during both freezing (blue, see Fig.\,\ref{fig:1}\,(b)\,\&\,(d)) and thawing (red, see Fig.\,\ref{fig:2}\,(c)\,\&\,(e)) as a function of time. The thawing time has been shifted for the curve to become continuous. The inset shows again the theoretical prediction.
(c) Sketch of a spherical particle interacting with an advancing solidification front \cite{meijer2024frozen}, indicating (among others) the particle-front distance $h(t)$, the distance between the particle's surface and the deformed front $d(\mathbf{x})$, and its radius $R$. The inset shows the quasi-stationary force balance, arising from disjoining pressure $\mathbf{F}_{\Pi}$ and viscous lubrication $\mathbf{F}_{\mathrm{vis}}$, to determine the particle velocity $u(t)$.  }
\label{fig:4}
\end{figure}

To answer this question we will first briefly review the basics of the theoretical model. 
A corresponding sketch including the definitions of the relevant quantities is shown in Fig.\,\ref{fig:4}\,(c).
For details on the precise underlying physics and the numerical implementation, we refer the reader to Ref.\cite{meijer2024frozen}. 
In summary, the aim of the model is to predict the velocity of the particle $u(t)$ as the solidification front approaches with velocity $V$.
The variation in the particle-front distance is then given by Eq.\,\ref{eq:dhdt} and the overall experienced displacement $l_{\mathrm{int}}$ follows after integrating the particle velocity in time following Eq.\,\ref{eq:lint}.

As the distance $d(\mathbf{x})$ between the surface of the particle and the deformed solid-liquid interface becomes smaller, intermolecular interactions between the particle, liquid, and solid lead to the presence of a Van der Waals disjoining pressure at the base of the particle.
As a consequence of this local increase in pressure, a thin liquid film, known as a premelted film \cite{wettlaufer2006premelting}, continuously separates the object from the solid, even during encapsulation.
It is assumed that these repulsive interactions are the main cause of the particle being pushed by the moving front.
The repelling force $\mathbf{F}_{\Pi}$ on the object can be obtained by integrating the disjoining pressure $\Pi = A/(6 \pi d(\mathbf{x})^3)$ over the particle's surface, with $A$ the Hamaker constant.
Opposing this force is a viscous friction force, $\mathbf{F}_{\mathrm{vis}}$, as the object moves through the melt, and the premelted liquid film at the particle's base must be replenished or drained.
Due to the assumed and justified quasi-stationarity \cite{meijer2024frozen} the particle velocity can then be obtained by a balance of these forces $\mathbf{F}_{\Pi} + \mathbf{F}_{\mathrm{vis}} = \mathbf{0}$.
For the latter, one also needs to know the precise shape of the solid-liquid interface, which at large distances is dictated by the earlier mentioned mismatch in the thermal conductivities between the particle an the melt, deflecting the interface towards or away from the object.
At the base of the object, a typical distance $d^* \sim R[A/(6 \pi R^2 \sigma_{\mathrm{sl}})]^{1/3}$ persists that is set by a balance between disjoining pressure and Laplace pressure \cite{meijer2024frozen}, \textit{i.e.}, $A/(6 \pi d^{*3}) \sim \sigma_{\mathrm{sl}}/R$, where $\sigma_{\mathrm{sl}}$ is the surface energy of the solidification front.
The force balance also gives rise to a typical velocity scale $\mathcal{W} = A/(\mu R^2)$, with $\mu$ the viscosity of the melt.
Now, by numerically solving the system of equations \citep{meijer2024frozen} one can model the interaction between a spherical object and an advancing solidification front, hence modelling the freezing process until the particle begins to be engulfed.
For the consecutive thawing process, we flip the sign of $V$ and let the front retract. 

In an effort to explain our experimental observations using the theoretical model, we include secondary effects not accounted for in the main model but introduced previously \cite{park2006encapsulation,meijer2024frozen}.
We include the effect of the volume change of water during phase change.
This gives rise to an extra force $\mathbf{F}_{\mathrm{vol}}$ on the object, that originates from alterations to the fluid flow at the base of the particle during its interaction with the moving front, leading to changes in the particle velocity.
Typically, $\mathbf{F}_{\mathrm{vol}} \propto V \left(\rho_l - \rho_s \right) / \rho_l$ and changes with the \textit{sign} of the velocity, as well as the extent of the change in density between the solid ($\rho_s$) and the liquid ($\rho_l$).
Incorporating this force into the model (see Ref.\cite{meijer2024frozen} for details) we obtain a slightly adjust model that allows for comparison to the experimental results.

For the PS particle interacting with the solidification front, the inset of Fig.\,\ref{fig:4}\,(a) shows the typical interaction we observe, represented by the theoretically determined $l_{\mathrm{int}}/R$.
Unsurprisingly, the freezing stage (blue data) is qualitatively nicely recovered and agrees with the experimental observation.
Remarkably, when letting the front retract (red data), the model is actually able to predict the additional push in the opposite direction of its motion, already observed experimentally.
Just as the particle has exited the ice, the particle-front distance has become very small and the repelling force $\mathbf{F}_{\Pi}$ becomes significant again. 
It therefore seems that the current model is capable of making valuable predictions on the displacement of PS particles in a strongly repulsive system, characterised by a large critical velocity (see Fig.\,\ref{fig:1}\,(e)). 
It should be noted that the model assumes that the particle and the front are sufficiently close together that the lubrication approximation in the thin liquid film remains valid.
As this gap increases in size, during the retraction of the front, this assumption will not hold any longer.
The model thus only provides valuable insights during the early stages of the front retraction and is currently not able to make predictions for the later stages of the thawing process.
Extending the theoretical framework to take this into account is beyond the scope of this paper.

This leaves us with the less repulsive oil-in-water emulsion, \textit{i.e.}, a system with a low critical velocity (see Fig.\,\ref{fig:1}\,(e)).
Here, we obtain a type of interaction that is depicted in the inset of Fig.\,\ref{fig:4}\,(b).
Now, we do see a displacement of the object towards its initial position over the course of the thawing process (red data), similar to what is observed experimentally.
In the case of droplets, a second potential candidate to cause this phenomena could be thermo-capillary (\textit{i.e.}, thermal Marangoni) forces due to flows at the free interface of the droplet, potentially leading to its migration \cite{park2006encapsulation,meijer2024frozen}.
However, since we do not observe any migration of the droplet as it is surrounded by the melt for over an extended period of time, we still assume it to be negligible.

\section{Conclusion \& outlook}
\label{sec:Conclusion}

To summarize, we have performed unidirectional freezing and thawing experiments on idealized model systems of oil-in-water emulsions and polystyrene (PS) particle suspensions.
The aim of the experiments was to study the interaction between an isolated spherical object with an advancing or retracting solidification front, and how the overall particle displacement is affected when exposed to a single freeze-thaw cycle.
For the case of the deformable oil droplets, we were also interested in the reversibility of the deformation dynamics during encapsulation into the ice as compared to its extraction during thawing.

We experimentally observed and delved into several distinct phenomena, ranging from no overall displacement of the objects for rapid freezing and thawing, an extended displacement away from its initial position throughout the entire freeze-thaw cycle for PS particles, and a return towards its initial position during thawing for oil droplets.
The latter two phenomena indicate a clear emergence of hysteresis in the overall displacement during a single freeze-thaw cycle. 
We rationalise the particle response by qualitatively comparing our experiments with our theoretical model of Ref.\cite{meijer2024frozen}, including a minor adjustment to account for volume changes during phase change, yielding a good agreement between the predictions of the model and the experimentally observed dynamics of the PS particles, for both the advancing front, for which the model was made, but remarkably also for the retracting front.
Additionally, the adjusted model also yields qualitatively promising results to rationalize the observed behaviour of the oil droplets during thawing.   
Finally, for the deformation dynamics of the droplets we find that, remarkably, this process is consistently reversible.

Although our experiments focus on the dynamics of single particles, exposed to only a single freeze-thaw cycle, we are confident that our current results already highlight the complexities of this process that need further investigations to be fully understood.
Our findings might inspire an extension of the theoretical framework to be applicable to extended thawing also.  
Lastly, and especially in the context of less dilute emulsions or suspensions, where interactions between the particles are inevitable, similar experiments as discussed here provide excellent research perspectives. 
These experiments could bridge the gap from idealised systems to those that are even more complex and realistic.

\section*{Conflicts of interest}
There are no conflicts to declare.

\section*{Data availability statement}
The materials underlying this study are provided in the Supplementary Material \cite{suppl}. Additional data and details supporting the findings are available from the corresponding authors upon reasonable request.

\section*{Acknowledgements}
The authors thank Gert-Wim Bruggert, Martin Bos, and Thomas Zijlstra for the technical support, as well as Prashanth Ramesh for valuable discussions.
The authors acknowledge the funding by Max Planck Center Twente and the Balzan Foundation.

\appendix
\subsection*{I - List of Supplementary Movies}

\begin{itemize}
\item \textbf{Movie 1:} Interaction between a silicone oil droplet of size $R \approx \SI{105}{\micro \meter}$ with a water-ice solidification front advancing at velocity $V \approx \SI{1}{\micro \meter \per \second}$.  

\item \textbf{Movie 2:} Interaction between a silicone oil droplet of size $R \approx \SI{105}{\micro \meter}$ with a water-ice solidification front advancing at velocity $V \approx \SI{0.7}{\micro \meter \per \second}$.  

\item \textbf{Movie 3:} Interaction between a silicone oil droplet of size $R \approx \SI{105}{\micro \meter}$ with a water-ice solidification front advancing at velocity $V \approx \SI{0.4}{\micro \meter \per \second}$.  

\item \textbf{Movie 4:} Interaction between a PS particle of size $R \approx \SI{20}{\micro \meter}$ with a water-ice solidification front retracting at velocity $V \approx \SI{-8.9}{\micro \meter \per \second}$.  

\item \textbf{Movie 5:} Interaction between a PS particle of size $R \approx \SI{70}{\micro \meter}$ with a water-ice solidification front retracting at velocity $V \approx \SI{-0.1}{\micro \meter \per \second}$.  

\item \textbf{Movie 6:} Interaction between a silicone oil droplet of size $R \approx \SI{105}{\micro \meter}$ with a water-ice solidification front retracting at velocity $V \approx \SI{-0.8}{\micro \meter \per \second}$.  

\end{itemize}

\subsection*{II - Droplet deformation dynamics}
\begin{figure}[b!]  % Figure over entire width
\includegraphics[width=\textwidth]{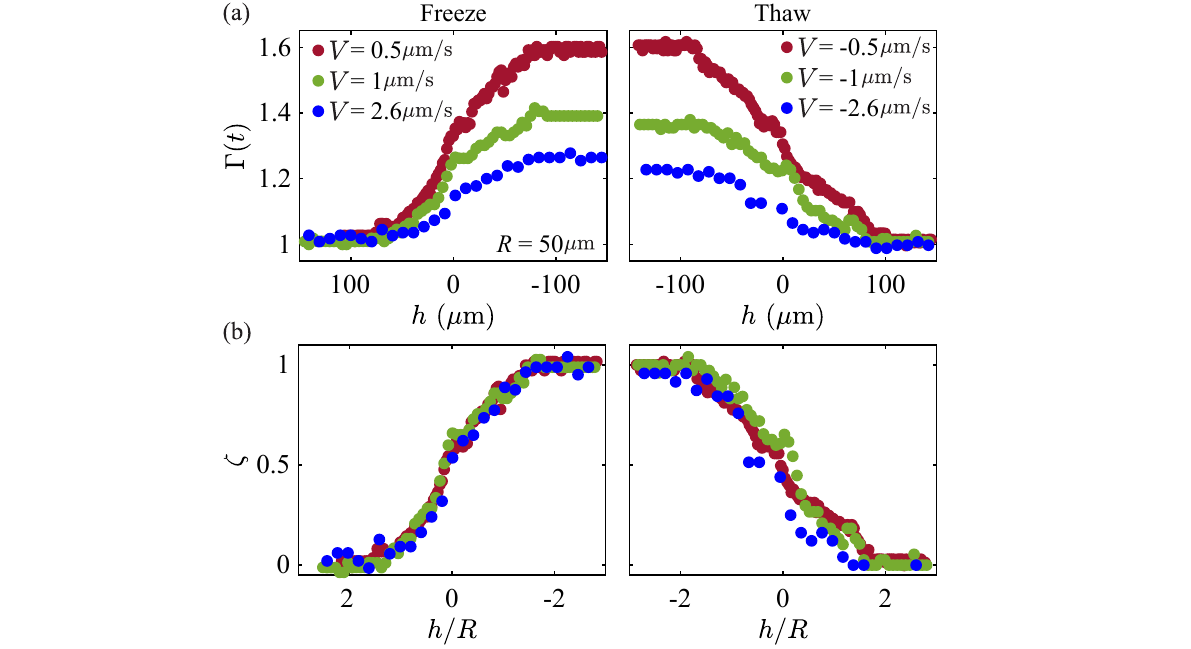}
\centering
\caption{(a) Droplet deformation profiles for the same droplet with $R = \SI{50}{\micro \meter}$ for three different freezing/thawing velocities. (b) Renormalised profiles where $\zeta = (\Gamma(t) - 1)/(\Gamma_{\infty} - 1)$.}
\label{fig:SI1}
\end{figure}

We perform consecutive freeze/thaw cycles on an individual oil droplet with $R = \SI{50}{\micro \m}$ at freezing/thawing velocities ranging from $V = \SI{0.5}{\micro \meter \per \second}$ to $V = \SI{4.1}{\micro \meter \per \second}$. Apart from the final deformation $\Gamma_{\infty}$ (see inset Fig.\,1(f) in the main text) we have also analysed the temporal evolution of $\Gamma(t)$ for cases at three different velocities. We measure $\Gamma(t)$ as a function of time during both freezing and thawing. By multiplying $t$ with the freezing/thawing velocity $V$ we can plot $\Gamma(t)$ as a function of particle-front distance $h(t)$ for better comparison, see Fig.\,\ref{fig:SI1}(a). We observe that faster freezing leads to less deformation of the droplet, characterisesd by the final deformation during freezing $\Gamma_{\infty}$. By renormalising $h(t)$ with the droplet radius and $\Gamma(t)$ with $\zeta = (\Gamma(t) - 1)/(\Gamma_{\infty} - 1)$ we find that the profiles collapse, for both freezing and thawing, see Fig.\,\ref{fig:SI1}\,(b). Lastly, the profiles for freezing and thawing are symmetric, yielding the same conclusion as in the main text that the deformation and re-formation of the droplet is a reversible process.

\subsection*{III - Velocity and size dependencies}

In section we highlight the velocity and size dependencies on the extra displacement for the solid particles and the pulling of the droplet by the front during thawing.

\subsubsection*{Solid (PS) particles}

\begin{figure}[t!]  % Figure over entire width
\includegraphics[width=\textwidth]{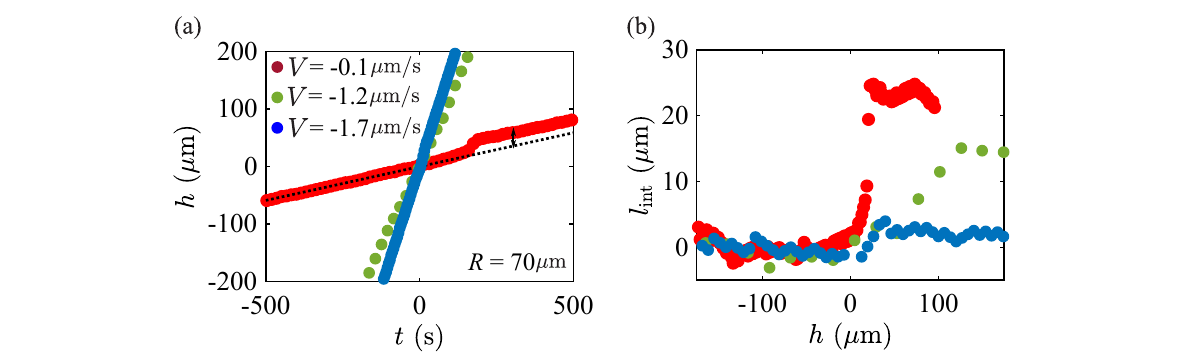}
\centering
\caption{(a) Particle-front distance $h(t)$ as a function of time during thawing of a PS particle with $R = \SI{70}{\micro \meter}$ at three different thawing velocities. The black arrow indicates how we determine the interaction length $l_{\mathrm{int}}$ (see also main text for definition). (b) Interaction length $l_{\mathrm{int}}$ as a function of $h(t)$ showing that the interaction weakens when the thawing velocity is increased. }
\label{fig:SI2}
\end{figure}

We again look at the particle-front distance $h(t)$ as a function of time for the PS particle with $R = \SI{70}{\micro \meter}$ discussed in the main text. The corresponding trajectory for $V = \SI{-0.1}{\micro \meter \per \second}$ is shown in Fig.\,2\,(d)\,\&\,(e) of the main text, as well as in Fig.\,\ref{fig:SI2}\,(a) below (red data set). In addition, we now show the trajectories for much larger velocities during the thawing process of a PS particle with the same radius. 
For the green data set the acquisition rate was lower hence the fewer points. \\
To amplify the interaction we plot the interaction length $l_{\mathrm{int}}$ as a function of particle-front distance $h(t)$ in  Fig.\,\ref{fig:SI2}\,(b). The definition of $l_{\mathrm{int}}$ is identical as described in the main text. We observe that the interaction between the solid particle and the front during thawing weakens when the thawing velocity is increased. The main physical mechanism underlying this additional push during thawing comes, we argue, from the disjoining pressure at the base of the particle during exiting the ice. For this repulsive force to be significant, the gap between the particle's surface and the ice interface should not increase too quickly, which is the case when thawing rapidly. The size of the particle (not investigated here) will also impact the extend of this additional push during thawing, as a larger particle experiences more drag and inertia, and a larger force is needed for its displacement.  

\subsubsection*{Oil droplets}
\begin{figure}[t!]  % Figure over entire width
\includegraphics[width=\textwidth]{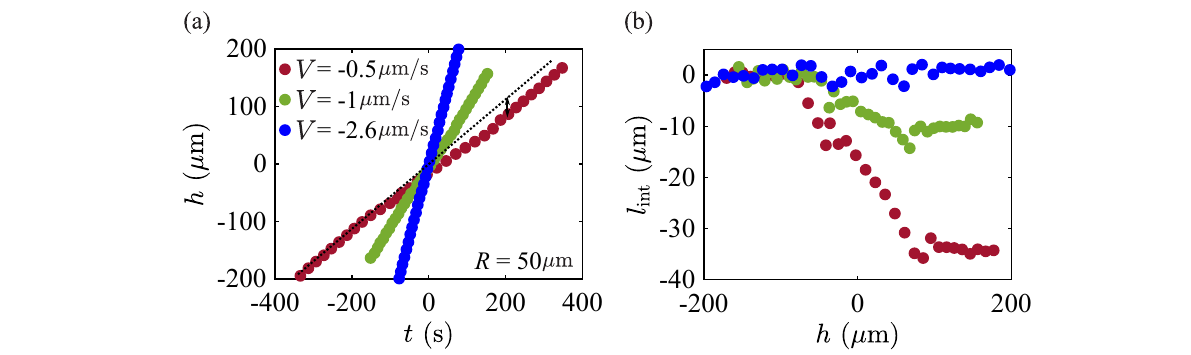}
\centering
\caption{(a) Particle-front distance $h(t)$ as a function of time during thawing of an oil droplet with $R = \SI{50}{\micro \meter}$ at three different thawing velocities. The black arrow indicates how we determine the interaction length $l_{\mathrm{int}}$ (see also main text for definition). (b) Interaction length $l_{\mathrm{int}}$ as a function of $h(t)$ showing that the interaction weakens when the thawing velocity is increased. }
\label{fig:SI3}
\end{figure}

We revisit the experiments performed in Sec.\,II and now look at the particle-front distance $h(t)$ as a function of time for the droplet with $R = \SI{50}{\micro \meter}$ and three different thawing velocities, see Fig.\,\ref{fig:SI3}\,(a). The definition of $t = 0$ is identical as in the main text and from the profiles we can see that as we go to faster thawing velocities, the interaction between the droplet and the front weakens. This feature is amplified when looking at the interaction length $l_{\mathrm{int}}$, which is again defined as discussed in the main text. The black arrow in Fig.\,\ref{fig:SI3}\,(a) indicates its meaning for the red set of points. \\
Fig.\,\ref{fig:SI3}\,(b) shows $l_{\mathrm{int}}$ for the three cases as a function of $h(t)$. It highlights the fact that the interaction between the droplet and the front during thawing weakens when the thawing velocity is increased. As for the PS particles, we argue that the rate of change of the gap distance between the particle's surface and the ice interface dictates the strength of the interaction.  \\
Comparing the final value of $l_{\mathrm{int}}$ of the red data set ($l_{\mathrm{int}} \approx \SI{35}{\micro \meter}$) to that of the droplet twice its size (see Fig.\,2\,(e), blue data set in the main text), shows that at roughly the same thawing velocity the larger droplet experiences a larger displacement during thawing. 

\bibliography{References}

%\clearpage
%\begin{widetext}
%
%\subsection*{I - List of Supplementary Movies}
%
%\begin{itemize}
%\item \textbf{Movie 1:} Interaction between a silicone oil droplet of size $R \approx \SI{105}{\micro \meter}$ with a water-ice solidification front advancing at velocity $V \approx \SI{1}{\micro \meter \per \second}$.  
%
%\item \textbf{Movie 2:} Interaction between a silicone oil droplet of size $R \approx \SI{105}{\micro \meter}$ with a water-ice solidification front advancing at velocity $V \approx \SI{0.7}{\micro \meter \per \second}$.  
%
%\item \textbf{Movie 3:} Interaction between a silicone oil droplet of size $R \approx \SI{105}{\micro \meter}$ with a water-ice solidification front advancing at velocity $V \approx \SI{0.4}{\micro \meter \per \second}$.  
%
%\item \textbf{Movie 4:} Interaction between a PS particle of size $R \approx \SI{20}{\micro \meter}$ with a water-ice solidification front retracting at velocity $V \approx \SI{-8.9}{\micro \meter \per \second}$.  
%
%\item \textbf{Movie 5:} Interaction between a PS particle of size $R \approx \SI{70}{\micro \meter}$ with a water-ice solidification front retracting at velocity $V \approx \SI{-0.1}{\micro \meter \per \second}$.  
%
%\item \textbf{Movie 6:} Interaction between a silicone oil droplet of size $R \approx \SI{105}{\micro \meter}$ with a water-ice solidification front retracting at velocity $V \approx \SI{-0.8}{\micro \meter \per \second}$.  
%
%\end{itemize}
%\end{widetext}
\end{document}